\documentclass[aps,twocolumn,nofootinbib,superscriptaddress,preprintnumbers,floatfix]{revtex4}

\usepackage{graphicx}
\usepackage{epsfig,amsmath,euscript}

\usepackage[hypertex]{hyperref}

\newcommand{\be}{\begin{equation}}
\newcommand{\ee}{\end{equation}}

\def\OMIT#1{{}}

\newcommand{\mcdot}{\!\cdot\!}

\newcommand{\bra}[1]{\left\langle #1\right\rvert}
\newcommand{\ket}[1]{\left\lvert #1\right\rangle}

\newcommand{\e}{\mathrm{e}}

\newcommand{\eq}[1]{Eq.~\eqref{#1}}
\newcommand{\vc}[1]{\boldsymbol{#1}}

\def\bnslash{{\bar{n}}\!\!\!\slash}

\newcommand{\GeV}{~\mathrm{GeV}}

\def\SCETG{${\rm SCET}_{\rm G}$}

\begin{document}

\title{Medium-induced parton splitting kernels from Soft Collinear Effective Theory \\[1ex] with Glauber gluons}
\author{Grigory Ovanesyan}
\affiliation{Los Alamos National Laboratory, Theoretical Division, MS B238, Los Alamos, NM 87545, USA}
\author{Ivan Vitev}
\affiliation{Los Alamos National Laboratory, Theoretical Division, MS B238, Los Alamos, NM 87545, USA}
\begin{abstract}
We derive the splitting kernels for partons produced in large $Q^2$
scattering processes that subsequently
traverse a region of strongly-interacting matter using a
recently-developed effective theory \SCETG.
We include all corrections beyond  the small-$x$ approximation,
consistent with the power counting
of \SCETG. We demonstrate how medium recoil, geometry and expansion
scenarios, and  phase space cuts
can be implemented numerically for phenomenological applications. For
the simplified case of infinite
transverse momentum kinematics and a uniform medium, we provide
closed-form analytic results that can
be used to validate the numerical simulations.
\end{abstract} 
\maketitle

\section{Introduction}\label{introsec}

The suppression in the production rate of energetic leading particles and particle correlations due 
to final-state  interactions in reactions with ultra-relativistic nuclei is among the best-known 
experimental discoveries at the Relativistic Heavy Ion Collider (RHIC)~\cite{Arsene:2004fa}-\cite{Adcox:2004mh}, 
and now at the Large Hadron Collider (LHC)~\cite{Aamodt:2010jd}-\cite{Dainese:2011vb}.
This jet quenching phenomenon also provides one of the strongest pieces of evidence for the
creation of dense strongly-interacting matter in such collisions~\cite{Gyulassy:2003mc}. 
Recent advances in understanding the modification of partons and parton showers in Quantum 
Chromodynamics (QCD) media have come from the inclusion of jets in the 
theoretical~\cite{Vitev:2009rd}-\cite{Neufeld:2011fh}  and 
experimental~\cite{Salur:2009vz}-\cite{Chatrchyan:2011ua} analyses. Jet observables are more
sensitive to the underlying theoretical model assumptions and the properties of the QCD medium
when compared to leading particle measurements~\cite{Vitev:2008rz}.

In recent years effective field theories (EFT) have become a powerful modern tool for jet physics. 
In particular, Soft Collinear Effective Theory~\cite{Bauer:2000ew, Bauer:2000yr, Bauer:2001ct, Bauer:2001yt} 
(SCET)  is an effective theory for QCD that describes the dynamics of highly energetic partons. 
It has been successfully  applied to improve the theoretical accuracy in the evaluation 
of  high energy cross sections at lepton~\cite{Fleming:2007xt}-\cite{Abbate:2010xh} and hadron 
colliders~\cite{Becher:2007ty}-\cite{Becher:2009th}.

The first step in constructing an effective theory for jets propagating in a QCD medium was done in 
Ref.~\cite{Idilbi:2008vm}, where the SCET Lagrangian was extended by adding a term that describes the 
interaction  of a quark jet with gluons that have momentum purely transverse to it, traditionally referred to in 
the literature as Glauber gluons. As an application of effective Lagrangian derived in~\cite{Idilbi:2008vm}, 
in Ref. \cite{D'Eramo:2010ak} the probability density of quark jet broadening~\cite{Gyulassy:2002yv,Qiu:2003pm} 
was re-derived as an expectation value of Wilson lines, which later the authors evaluate using 
AdS/CFT correspondence.  In Ref.~\cite{Ovanesyan:2011xy} the Yang-Mills part of the 
collinear SCET Lagrangian was coupled to Glauber gluons, which allowed to perform calculations 
for parton splitting processes in the medium. The resulting effective theory was called $\text{\SCETG}$, where ``G" 
stands for Glauber gluons. In that paper a detailed connection was made between calculations in 
$\text{\SCETG}$ and the evaluation of jet broadening~\cite{Gyulassy:2002yv,Qiu:2003pm}  and 
medium-induced quark energy loss in the Gyulassy-Levai-Vitev approach~\cite{Gyulassy:2002yv,Vitev:2007ve}.  
The gauge invariance of the physics results was explicitly demonstrated for three 
different gauge choices. One of the medium-induced splittings, namely $q\rightarrow qg$ has been calculated in Ref.~\cite{Ovanesyan:2011xy}
beyond the soft emitted gluon approximation. There are three additional splittings: $g\rightarrow gg, g\rightarrow q\bar{q}$ and $q\rightarrow gq$. To complete the derivation of all medium-induced  branching processes
without the assumption of a soft final-state parton and to understand the correction that arise from 
the finite parton scattering kinematics, branching kinematics, and recoil of the constituents of the
QCD medium is the main goal of this Letter.

The rest of this Letter is organized as follows: in section~\ref{framework} we discuss the theoretical
framework $\text{\SCETG}$ for our calculation and demonstrate how the vacuum Altarelli-Parisi splitting kernels 
can  be derived in Soft Collinear Effective Theory. The derivation of the full splitting 
kernels for in-medium jet production with final-state interactions is discussed in section~\ref{medsplit}.
We elucidate the relation to early soft gluon approximation results and provide analytic formulas 
for simplified kinematics and medium geometry scenarios. Numerical control over the newly-derived 
medium-induced  splitting intensities is demonstrated in section~\ref{numericssec}. In this section we 
also quantify  the effects of  large-$x$ corrections, finite kinematics, and medium recoil.   
A brief summary and outlook is presented in section~\ref{conclusions}.

\section{Theoretical Framework}
\label{framework}

An effective theory, well-suited to describing the propagation of jets in the medium, has been recently 
developed in Ref.~\cite{Ovanesyan:2011xy}. The Lagrangian of this EFT is given by the sum of the SCET 
Lagrangian~\cite{Bauer:2000ew, Bauer:2000yr, Bauer:2001ct, Bauer:2001yt} and a term that specifies 
the interactions of collinear partons in QCD matter: 
\begin{eqnarray}
&& \!\!\!\!\!\!  \mathcal{L}_{\text{\SCETG}}(\xi_n, A_n, A_G)=\mathcal{L}_{\text{SCET}}(\xi_n, A_n)+
\mathcal{L}_{\text{G}}\left(\xi_n, A_n, A_G\right),\nonumber\\
 && \!\!\!\!\!\! \mathcal{L}_{\text{G}}\left(\xi_n, A_n, A_G\right)=\sum_{p,p'}\e^{-i(p-p')x}\Big(\bar{\xi}_{n,{p'}} 
\Gamma^{\mu,a}_{\rm qqA_G}\frac{\bnslash}{2}\xi_{n,p}\nonumber\\
&& \!\!\!\!\!\! \qquad\qquad\qquad-i \Gamma^{\mu\nu\lambda,abc}_{\rm ggA_G}\,
\left({A}^{b}_{n, p'}\Big)_{\nu}\left({A}^{c}_{n, p}\right)_{\lambda}\right)\, 
A_{{\rm G}\, \mu, a}(x)\label{LGdef0}\, . \qquad\nonumber\\
\end{eqnarray}
In Ref.~\cite{Ovanesyan:2011xy} the vertexes $\Gamma^{\mu,a}_{\rm qqA_G}, \Gamma^{\mu\nu\lambda,abc}_{\rm ggA_G}$ have 
been derived for three types of gauge-fixing conditions: covariant, light-cone and hybrid gauges. In the first 
case we gauge-fix both the physical collinear gluons as well as the Glauber gluons in the covariant gauge. 
The second choice corresponds to gauge-fixing both fields using the light-cone gauge. The third choice, which 
appears to be the most convenient from the practical point of view, corresponds to a light-cone gauge for 
collinear gluons and a covariant gauge for the Glauber gluons. This is a legitimate choice from effective theory point of view, since we are allowed to gauge-fix separate gauge sectors independently. Another way of justifying this gauge choice is factorization between the splitting and the elastic scattering.
In this hybrid case both the collinear Wilson line $W$ 
and the transverse gauge link $T$  \cite{Ji:2002aa,Idilbi:2010im, GarciaEchevarria:2011md} vanish.  Gauge invariance of the physics results for the in-medium elastic 
scattering and  radiative  energy loss was demonstrated in~\cite{Ovanesyan:2011xy}, providing a cross-check 
on the  approach  and the newly-derived Feynman rules.  
It is interesting to note that the same effective theory $\text{\SCETG}$ 
is relevant for describing the Drell-Yan process, as shown in Ref.~\cite{Bauer:2010cc}.

We start from amplitudes for the parton  splitting processes:
\begin{eqnarray}
A_{q\rightarrow q g}&=&\bra{q(p) g(k)}T\,\e^{iS}\,\bar{\chi}_n(x_0)\ket{q(p_0)},\label{A1def}\\
A_{g\rightarrow gg}&=&\bra{g(p) g(k)}T\,\e^{iS}\,\mathcal{B}^{\lambda c}(x_0)\ket{g(p_0)},\label{A3def}\\
A_{g\rightarrow q \bar{q}}&=&\bra{q(p) \bar{q}(k)}T\,\e^{iS}\,\mathcal{B}^{\lambda c}(x_0)\ket{g(p_0)},\label{A2def}\\
A_{q\rightarrow g q}&=&\bra{g(p) q(k)}T\,\e^{iS}\,\bar{\chi}_n(x_0)\ket{q(p_0)},\label{A4def}
\end{eqnarray}
where $\chi, \mathcal{B}$ are collinear gauge invariant SCET fields~\cite{Arnesen:2005nk, Bauer:2008qu}
and the momentum  
four-vectors, such as $p_0=p+k$, are parametrized in the standard way, consistent with energy 
momentum conservation and the on-shell condition $p^2=k^2=0$: 
\begin{eqnarray}
p_0&=&\left[p_0^+, \frac{{\bf{k}}_{\perp}^2}{x(1-x)p_0^+}, {\bf{0}}_{\perp}\right],\\
p&=&\left[(1-x)p_0^+, \frac{{\bf{k}}_{\perp}^2}{(1-x)p_0^+}, -{\bf{k}}_{\perp}\right],\\
k&=&\left[x p_0^+, \frac{{\bf{k}}_{\perp}^2}{x p_0^+}, {\bf{k}}_{\perp}\right].
\end{eqnarray}
We use square brackets to indicate the light-cone notation, which we define for arbitrary four-vector $q$ in the following way: $q\equiv\left[q^{+},q^{-},\vc{q}_{\perp}\right]=\left[\bar{n}\mcdot q, n\mcdot q, \vc{q}_{\perp}\right]$ and $n^{\mu}=\left(1,0,0,1\right), \bar{n}^{\mu}=\left(1,0,0,-1\right)$. The action in \eq{A1def}-\eq{A4def} is given by Lagrangian of $\text{\SCETG}$:
\begin{eqnarray}
S=i\int \text{d}^{4}x\, \mathcal{L}_{\text{\SCETG}}.
\end{eqnarray}
The Lagrangian of $\text{\SCETG}$~\cite{Idilbi:2008vm, Ovanesyan:2011xy} is given in \eq{LGdef0} and it 
evolves the created jet and 
describes the parton splitting processes and the interaction of the parton shower in the medium.
The amplitude with $q\bar{q} \leftrightarrow \bar{q}q$ is not shown explicitly.

Restricting ourselves to the SCET Lagrangian without Glauber gluons, we first verify that at 
tree level we  recover the  Altarelli-Parisi splitting kernels~\cite{Altarelli:1977zs}, which have 
been originally calculated in full QCD: 
\begin{eqnarray}
\label{qqg}
   \left(\frac{dN}{dx \,d^2{\bf{k}}}_{\perp}\right)_{q\rightarrow qg}&=&
\frac{\alpha_s}{2\pi^2} C_F \frac{1+(1-x)^2}{x}\frac{1}{{\bf{k}}_{\perp}^2},\\
\label{ggg}
    \left(\frac{dN}{dx \,d^2{\bf{k}}}_{\perp}\right)_{g\rightarrow gg}&=& \frac{\alpha_s}{2\pi^2}
  2 C_A \Big(\frac{1-x}{x}+\frac{x}{1-x}  \nonumber\\
&&   \qquad\qquad +x(1-x) \Big)\frac{1}{{\bf{k}}_{\perp}^2},\\
\label{gqq}
    \left(\frac{dN}{dx \,d^2{\bf{k}}}_{\perp}\right)_{g\rightarrow q\bar{q}}&=
&  \frac{\alpha_s}{2\pi^2} T_R\  \left( x^2+(1-x)^2 \right)\frac{1}{{\bf{k}}_{\perp}^2},\\
   \left(\frac{dN}{dx \,d^2{\bf{k}}}_{\perp}\right)_{q\rightarrow gq}&=&
\label{qgq}
  \left(\frac{dN}{dx \,d^2{\bf{k}}}_{\perp}\right)_{q\rightarrow qg} (x\rightarrow 1-x).\nonumber\\
 \end{eqnarray}
We note that we are interested in real splitting processes away from the singular end 
points $x=0$ and $x=1$. In all expressions above the transverse momentum $\vc{k}_{\perp}$ and the 
lightcone momentum 
fraction $x  =k^+/p_0^+ = k^+/(p^+ + k^+)$ are for the second final-state parton. The parent parton has no net 
transverse  momentum and  $\vc{k}_{\perp}=-\vc{p}_{\perp}$. Note that Eq.~(\ref{qqg}) and  Eq.~(\ref{qgq}) 
are  interchangeable  under  $x\rightarrow 1-x$, whereas Eq.~(\ref{ggg}) and  Eq.~(\ref{gqq}) are symmetric 
under this substitution. The same symmetries hold for the medium-induced splittings that we derive in 
section~\ref{medsplit}.

In this paper we use the following terminology: the double differential distribution 
$dN/dx d^2{\bf{k}}_{\perp}$ we call a splitting kernel, $x dN/dx$ we call a splitting intensity and ${dN}/{dx}$ we call a differential
emitted parton number distribution. This terminology applies to both vacuum and medium-induced splittings. The  $x-$dependent part
of the vacuum splitting kernel we call a splitting function. Since the medium-induced kernel has a more complicated $\vc{k}_{\perp}, x$ correlation
structure compared to the simple factorized form in Eq. (10) -- Eq. (13) we avoid definition of a similar term in the medium.

\begin{figure}[!t]
\includegraphics[width=255pt]{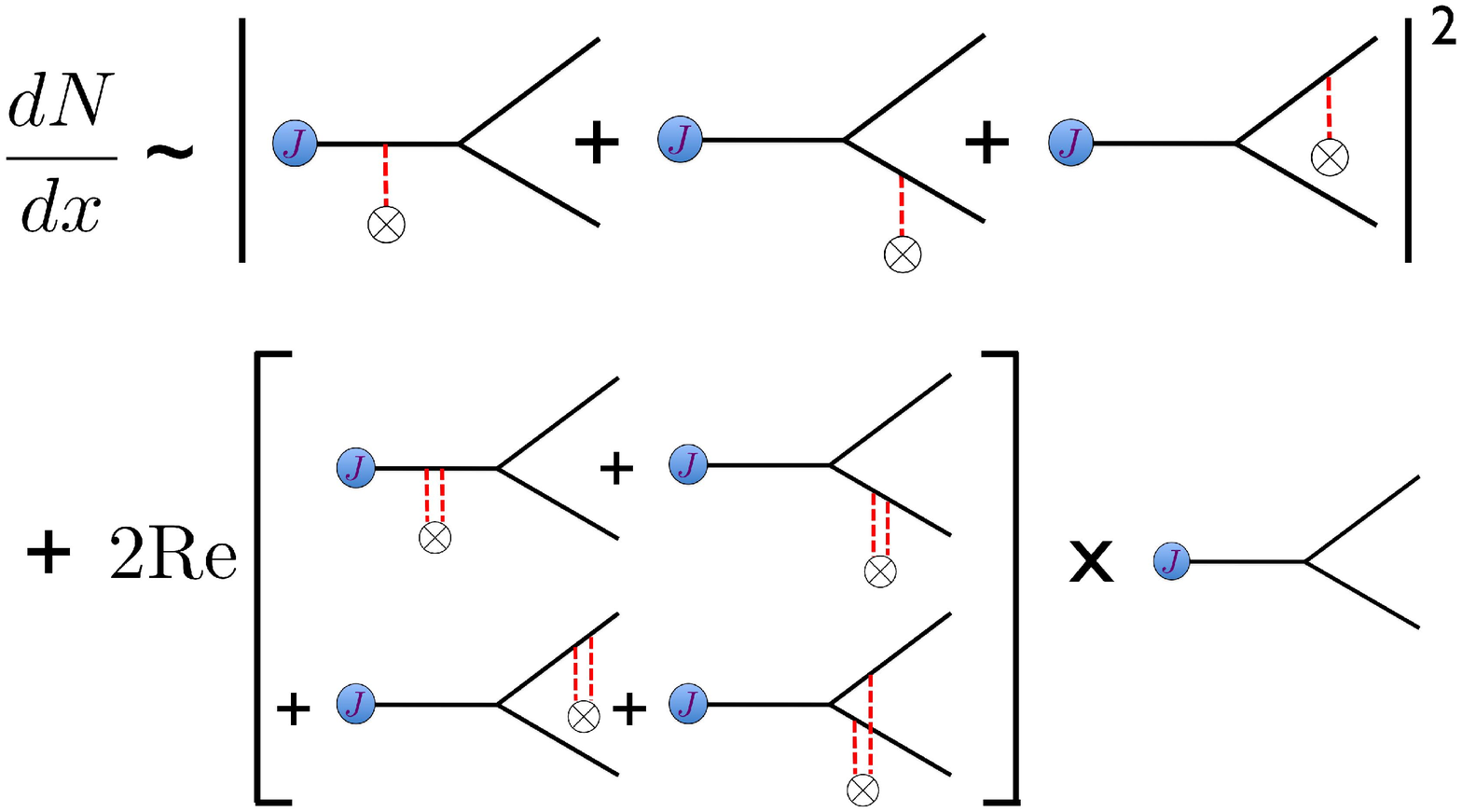}
\caption{Feynman diagrams contributing to medium-induced splittings at first order in opacity. 
Red lines corresponds to Glauber gluons. The kinematics and topology are common to all splitting processes:
$q\rightarrow qg$, $g\rightarrow gg$, $g\rightarrow q \bar{q}$, $q\rightarrow gq$.}
\label{fig:diagrams}
\end{figure}

\section{ Medium-induced parton splittings }\label{medsplit}
To describe the collisional and radiative processes for partons propagating in QCD matter,  
both single and double Glauber gluon exchanges between the jets and the constituents of the medium must be 
considered~\cite{Gyulassy:2000er,Vitev:2007ve,Ovanesyan:2011xy}. The calculation to first order in opacity, 
which takes into  account the contribution from the splitting induced by the interactions along the trajectory 
of the parent parton and the dominant interference with the splitting induced by the large $Q^2$ process, 
is illustrated in figure~\ref{fig:diagrams}. We do not specify the parent and daughter parton flavors since 
the  topology and kinematics are the same for the splitting processes enumerated in 
Eqs.~(\ref{qqg}) - (\ref{qgq}).
Consequently, all results can be expressed in terms of universal transverse momentum vectors  
$\vc{A}_{\perp}, \vc{B}_{\perp}, \vc{C}_{\perp}, \vc{D}_{\perp}$ 
and interference phases $\Omega_1,...,\Omega_5$, defined in~\cite{Ovanesyan:2011xy}:
\begin{eqnarray}
&&\vc{A}_{\perp}=\vc{k}_{\perp},\,\, \vc{B}_{\perp}=\vc{k}_{\perp} + x \vc{q}_{\perp} , \,\,
\vc{C}_{\perp}=\vc{k}_{\perp} -  (1-x)\vc{q}_{\perp},\nonumber\\[1ex]
&& \vc{D}_{\perp}=\vc{k}_{\perp}-\vc{q}_{\perp},\,\, \\
&&\Omega_1-\Omega_2=\frac{\vc{B}_{\perp}^2}{p_0^+ x(1-x)}, \,\Omega_1-\Omega_3=\frac{\vc{C}_{\perp}^2}{p_0^+x(1-x)},
\nonumber\\
&& \Omega_2-\Omega_3=\frac{\vc{C}_{\perp}^2-\vc{B}_{\perp}^2}{p_0^+x(1-x)}, \,\,
\Omega_4=\frac{\vc{A}_{\perp}^2}{p_0^+x(1-x)}, \nonumber \\
&& \Omega_5=\frac{\vc{A}_{\perp}^2-\vc{D}_{\perp}^2}{p_0^+x(1-x)},
\end{eqnarray}
where $ p_0^+ = p^+ + k^+ $ and the parent parton has no net transverse momentum. 

For completeness, we first present below the result for the $q\rightarrow q g$ splitting, 
calculated  in~\cite{Ovanesyan:2011xy} and shown to be gauge invariant:
\begin{widetext}
\begin{eqnarray}
 &&  \left( \frac{dN}{ dxd^2\vc{k}_{\perp} }\right)_{q\rightarrow qg}  =  \frac{\alpha_s}{2\pi^2}
C_F  \frac{1+(1-x)^2}{x}  
\int \frac{d\Delta z}{\lambda_g(z)}  
\int d^2{\bf q}_\perp  \frac{1}{\sigma_{el}} \frac{d\sigma_{el}^{\; {\rm medium}}}{d^2 {\bf q}_\perp} \; 
 \Bigg[  \frac{\vc{B}_{\perp}}{\vc{B}_{\perp}^2} \mcdot \left( \frac{\vc{B}_{\perp}}{\vc{B}_{\perp}^2}  -  \frac{\vc{C}_{\perp}}{\vc{C}_{\perp}^2}   \right)
   \nonumber \\
&&  \qquad \qquad  
  \times \big( 1-\cos[(\Omega_1 -\Omega_2)\Delta z] \big)  + \frac{\vc{C}_{\perp}}{\vc{C}_{\perp}^2} \mcdot \left( 2 \frac{\vc{C}_{\perp}}{\vc{C}_{\perp}^2}   
-    \frac{\vc{A}_{\perp}}{\vc{A}_{\perp}^2} - \frac{\vc{B}_{\perp}}{\vc{B}_{\perp}^2}  \right) \big(1- \cos[(\Omega_1 -\Omega_3)\Delta z] \big) \nonumber \\  
&&
   \qquad \qquad    + \frac{\vc{B}_{\perp}}{\vc{B}_{\perp}^2} \mcdot \frac{\vc{C}_{\perp}}{\vc{C}_{\perp}^2} 
\big( 1 -  \cos[(\Omega_2 -\Omega_3)\Delta z] \big)  
+ \frac{\vc{A}_{\perp}}{\vc{A}_{\perp}^2} \mcdot \left( \frac{\vc{D}_{\perp}}{\vc{D}_{\perp}^2} - \frac{\vc{A}_{\perp}}{\vc{A}_{\perp}^2} \right) 
\big(1-\cos[\Omega_4\Delta z]\big)  \nonumber \\
&& \qquad \qquad  -\frac{\vc{A}_{\perp}}{\vc{A}_{\perp}^2} \mcdot \frac{\vc{D}_{\perp}}{\vc{D}_{\perp}^2}\big(1-\cos[\Omega_5\Delta z]\big)   
+  \frac{1}{N_c^2}  \frac{\vc{B}_{\perp}}{\vc{B}_{\perp}^2} \mcdot  \left( \frac{\vc{A}_{\perp}}{\vc{A}_{\perp}^2}  -   
\frac{\vc{B}_{\perp}}{\vc{B}_{\perp}^2}      \right)
\big( 1-\cos[(\Omega_1 -\Omega_2)\Delta z] \big)   \Bigg] \, ,           
\label{CohRadSX1} 
\end{eqnarray} 
\end{widetext}
 where $\lambda_g(z)$ is the scattering length of a gluon in the medium and $\left(1/\sigma_{el}\right) \,{d\sigma_{el}^{\; {\rm medium}}}/{d^2 {\bf q}_\perp}$ stands for normalized elastic scattering cross section of a parton in the medium. Even though this quantity varies when parton is a quark or a gluon, in the 
 high energy limit, when the $t-$ channel dominates the elastic scattering, this normalized cross section does not change significantly.
 
 Using the Feynman rules of~\text{\SCETG} in the hybrid gauge and the Feynman  diagrams exactly 
analogous to the case of $q\rightarrow q g$ splitting considered in~\cite{Ovanesyan:2011xy} and shown in figure \ref{fig:diagrams}, we 
derive the remaining parton splittings in the medium. The calculations are non-trivial and facilitated by 
intermediate results in~\cite{Ovanesyan:2011xy}. As discussed in section~\ref{framework}, the medium-induced 
splitting for $q \rightarrow gq$ can be obtained from Eq.~(\ref{CohRadSX1}) with the substitution 
$x\rightarrow 1-x$. Here, we skip the explicit expression for brevity.  The remaining  two splittings from a 
parent gluon are as follows:   
\begin{widetext}
\begin{eqnarray}
&&   \left( \frac{dN}{ dxd^2\vc{k}_{\perp} }\right)_{ \left\{ \begin{array}{c}   g \rightarrow gg\\     g\rightarrow q\bar{q}  \end{array} \right\} } 
= 
 \left\{ \begin{array}{c}     \frac{\alpha_s}{2\pi^2} \, 2 C_A \left(\frac{x}{1-x}+\frac{1-x}{x}+x(1-x) \right)  \\[1ex]
                           \frac{\alpha_s}{2\pi^2}  T_R \left( x^2+(1-x)^2 \right)  \end{array} \right\}
\int {d\Delta z}   \left\{ \begin{array}{c}    \frac{1}{\lambda_g(z)} \\[1ex]  \frac{1}{\lambda_q(z)}    \end{array} \right\}
\int d^2{\bf q}_\perp  \frac{1}{\sigma_{el}} \frac{d\sigma_{el}^{\; {\rm medium}}}{d^2 {\bf q}_\perp} \;  \nonumber \\
&& \qquad \qquad
\times \Bigg[  2\, \frac{\vc{B}_{\perp}}{\vc{B}_{\perp}^2} \mcdot \left(\frac{\vc{B}_{\perp}}{\vc{B}_{\perp}^2}-\frac{\vc{A}_{\perp}}{\vc{A}_{\perp}^2}\right) \big( 1-\cos[(\Omega_1 -\Omega_2)\Delta z]  \big)   
+2\, \frac{\vc{C}_{\perp}}{\vc{C}_{\perp}^2} \mcdot \left(\frac{\vc{C}_{\perp}}{\vc{C}_{\perp}^2}-\frac{\vc{A}_{\perp}}{\vc{A}_{\perp}^2}\right) \big( 1-\cos[(\Omega_1 -\Omega_3)\Delta z]  \big)
\nonumber \\
 &&\qquad \qquad +  \left\{ \begin{array}{c}   - \frac{1}{2}   \\[1ex] \frac{1}{N_c^2-1} \end{array} \right\}
\Bigg(2 \frac{\vc{B}_{\perp}}{\vc{B}_{\perp}^2}\mcdot\left(\frac{\vc{C}_{\perp}}{\vc{C}_{\perp}^2}-\frac{\vc{A}_{\perp}}{\vc{A}_{\perp}^2}\right)\big(1-\cos[(\Omega_1-\Omega_2)\Delta z ]\big)
\nonumber\\
 &&\qquad \qquad  +2\,\frac{\vc{C}_{\perp}}{\vc{C}_{\perp}^2}\mcdot\left(\frac{\vc{B}_{\perp}}{\vc{B}_{\perp}^2}-\frac{\vc{A}_{\perp}}{\vc{A}_{\perp}^2}\right)\big(1-\cos[(\Omega_1-\Omega_3)\Delta z]\big)-2\,\frac{\vc{C}_{\perp}}{\vc{C}_{\perp}^2}\mcdot \frac{\vc{B}_{\perp}}{\vc{B}_{\perp}^2}\big(1-\cos[(\Omega_2-\Omega_3)\Delta z]\big)
\nonumber\\
 &&\qquad \qquad +2\,\frac{\vc{A}_{\perp}}{\vc{A}_{\perp}^2}\mcdot\left(\frac{\vc{A}_{\perp}}{\vc{A}_{\perp}^2}-\frac{\vc{D}_{\perp}}{\vc{D}_{\perp}^2}\right)\big(1-\cos[\Omega_4\Delta z]\big)+2\,\frac{\vc{A}_{\perp}}{\vc{A}_{\perp}^2}\mcdot \frac{\vc{D}_{\perp}}{\vc{D}_{\perp}^2}\big(1-\cos[\Omega_5\Delta z]\big)\Bigg) \Bigg] \, ,      
\label{CohRadSX2} 
\end{eqnarray} 
\end{widetext}
where $\lambda_q(z)$ is the scattering length of a quark in the medium and same comment applies to the quantity $\left(1/\sigma_{el}\right) \,{d\sigma_{el}^{\; {\rm medium}}}/{d^2 {\bf q}_\perp}$  as after \eq{CohRadSX1}. Note that up to the overall vacuum-like splitting functions and color factors reflected both in the mean free 
paths (quark versus gluon)  and the corrections relevant beyond the small-$x$ approximation, the structure 
of the answers is very similar. The symmetry of $g\rightarrow gg, g\rightarrow q\bar{q}$ splitting 
kernels under $x \rightarrow 1-x $ is most easily verified explicitly by realizing that the parton scattering 
cross section in the medium is invariant under  $\vc{q}_\perp \rightarrow   -\vc{q}_\perp$.

The basic features of the medium-induced parton splitting kernels are:
\begin{itemize}
 \item In QCD, for parent quark they factorize from the hard scattering cross section up to a standard integral convolution \cite{Ovanesyan:2011xy}.
For parent gluons non-trivial spin correlation are present analogous to the vacuum case~\cite{Catani:1998nv}. 
\item They  are proportional to their vacuum Altarelli-Parisi splitting functions~\cite{Altarelli:1977zs}. 
\item  The in-medium splittings  are gauge-invariant, as they should be, since the underlying 
jet production process itself is gauge-invariant~\cite{Ovanesyan:2011xy}. 
\item The splitting kernels  depend on the properties of the QCD matter and vanish 
when the size or density of the medium vanish. The functions derived here are only valid for final-state 
interactions~\cite{Vitev:2007ve}. 
\end{itemize}

It is instructive to verify that in the small-$x$ limit  only two of the four medium-induced splitting
intensities survive and this allows for the standard energy loss interpretation of jet quenching:     
\begin{eqnarray}
 && \!\!\!\!\!\!\!  x \left( \frac{dN}{ dx}\right)_{ \left\{ \begin{array}{c}   q \rightarrow qg \\    
 g\rightarrow gg  \\  g \rightarrow q\bar{q} \\ q \rightarrow gq\\   \end{array} \right\} }  =  \frac{\alpha_s}{\pi^2} 
  \left\{ \begin{array}{c}   C_F[ 1+ {\cal O}(x) ] \\  C_A[ 1+ {\cal O}(x) ]  \\
  T_R [0 + \frac{x}{2} + {\cal O}(x^2)]  \\    C_F [0 + \frac{x}{2} + {\cal O}(x^2)]
  \end{array} \right\} \nonumber \\
&&   \qquad  \times    \int {d\Delta z}   \left\{ \begin{array}{c}    \frac{1}{\lambda_g(z)} \\[1ex]  \frac{1}{\lambda_g(z)}  \\
 \frac{1}{\lambda_q(z)}  \\  \frac{1}{\lambda_q(z)}    \end{array} \right\}
\int d^2{\bf k}_\perp d^2{\bf q}_\perp  \frac{1}{\sigma_{el}} \frac{d\sigma_{el}^{\; {\rm medium}}}{d^2 {\bf q}_\perp} \;  \nonumber \\
&&  \qquad   \times \frac{2 \vc{k}_{\perp}  \cdot \vc{q}_{\perp} }{\vc{k}_{\perp}^2 (\vc{k}_{\perp}-\vc{q}_{\perp})^2}
   \left [ 1-\cos \frac{   (\vc{k}_{\perp}-\vc{q}_{\perp})^2}{xp^+_0} \Delta z \right].
\label{smallx}
\end{eqnarray} 
In this limit the interference structure for all medium-induced splitting intensities is the same. 
Furthermore, in the small-$x$ limit 
the last two splittings are suppressed (${\mathcal {O}}(x)$) relative to the first two. 
We keep the first correction for numerical
comparison only. The color structure for the in-medium interactions also simplifies in this 
limit and is determined by the flavor of the small-$x$ parton in the final state. Specifically, 
the first two in-medium splittings are proportional to 
$1/\lambda_g$ and the second two   are proportional to  $1/\lambda_q$. In deriving these results, 
we have used relation: $\lambda_q/\lambda_g = C_A/C_F$, which follows from the leading order perturbation theory approximation. As expected, in the small-$x$ emission limit our results 
coincide exactly with the 
intensity derived (or neglected when the leading term is 0) in~\cite{Gyulassy:2000er}.

In section~\ref{numericssec} we will study numerically the in-medium splittings  derived here with an 
emphasis on going beyond the traditional small-$x$ approximation and on including medium recoil. The remaining 
part of the current section is devoted to deriving analytic formulas for the inclusive splitting intensity 
$x (dN/dx) $ under certain idealized assumptions. This will, in turn, allow us to obtain fully analytic 
formulas that can be used to benchmark the realistic numerical calculation.

A useful starting point for integrating the splitting kernels analytically over the transverse momenta  
is the following master formula:  
\begin{eqnarray}
&&\int d^2\vc{k}_{\perp}d^2\vc{q}_{\perp}   \frac{1}{\sigma_{el}} \frac{d\sigma_{el}}{d^2 {\bf q}_\perp}
\frac{2 \vc{k}_{\perp}  \cdot \vc{q}_{\perp} }{\vc{k}_{\perp}^2 (\vc{k}_{\perp}-\vc{q}_{\perp})^2}
\nonumber\\ 
&& \qquad \qquad \times  \left [ 1-\cos(\alpha(\vc{k}_{\perp}-\vc{q}_{\perp})^2) \right]  =f[\alpha\mu^2], \qquad
\label{masterformula}
\end{eqnarray}
where a specific form for elastic cross section was assumed as explained below and shown in \eq{infsigmael}, and function $f[x]$ equals:
\begin{eqnarray}
f[x]&=& 2\pi\Big[ \gamma_{\text{E}}+\ln(x)+\frac{\pi}{2}\sin(x)-\cos(x)\text{Ci}(x)  \nonumber \\
&& -\sin(x)\text{Si}(x) \Big].
\label{comfundef}
\end{eqnarray}
Two assumptions are already made at this level. First, we took the limits of integration on 
$\vc{k}_{\perp}, \vc{q}_{\perp}$ to infinity. In reality, phase space cuts affect the cross section and we 
study this effect numerically in the next section. Second, we neglected the recoil effect in the medium. 
In that approximation the normalized cross section equals:  
\begin{eqnarray}
\frac{1}{\sigma_{el}} \frac{d\sigma_{el}}{d^2 {\bf q}_\perp}
=\frac{\mu^2}{\pi(\vc{q}_{\perp}^2+\mu^2)^2}.\label{infsigmael}
\end{eqnarray}
The effects of finite medium recoil are also studied in the next section numerically. It turns out that 
using \eq{masterformula} it is possible to calculate $\vc{k}_{\perp}, \vc{q}_{\perp}$ integrals in all 
in-medium splittings \eq{CohRadSX1}-\eq{CohRadSX2}.  
The result is rather compact and can be expressed in terms of the function $f[x]$ defined in \eq{comfundef}:
\begin{widetext}
\begin{eqnarray}
&&x\left(\frac{dN}{dx}\right)^{\infty}_{q\rightarrow qg}= x\frac{\alpha_s}{2\pi^2} C_F 
\frac{ 1 + (1-x)^2}{x} \int\frac{d\Delta z}{\lambda_g(z)}
\frac{f[\beta]+f[\beta(1-x)^2]-\frac{1}{N_c^2}f[\beta x^2]}{2}, \label{split1analytics}\\
&&x\left(\frac{dN}{dx}\right)^{\infty}_{g\rightarrow gg}= x\frac{\alpha_s}{2\pi^2} 
2C_A \left(\frac{x}{1-x}+\frac{1-x}{x}+x(1-x)\right)
\int\frac{d\Delta z}{\lambda_g(z)}\frac{f[\beta x^2]+f[\beta(1-x)^2]+f[\beta]}{2},\\
&&x\left(\frac{dN}{dx}\right)^{\infty}_{g\rightarrow q\bar{q}}= x\frac{\alpha_s}
{2 \pi^2} T_R  \left(x^2+(1-x)^2\right)\int\frac{d\Delta z}{\lambda_q(z)} 
 \left[{\frac{N_c^2}{N_c^2-1}\left(f[\beta x^2]+f[\beta(1-x)^2]\right)-\frac{1}{N_c^2-1}f[\beta]}\right],\\
 &&x\left(\frac{dN}{dx}\right)^{\infty}_{q\rightarrow gq}=x \left(\frac{dN}{dx}\right)^{\infty}_{q\rightarrow qg}(x\rightarrow 1-x), \label{split4analytics}
\end{eqnarray}
\end{widetext}
where the superscript $\infty$ stands for infinite limits of integrations for $\vc{k}_{\perp}, \vc{q}_{\perp}$, and:
\begin{equation}
 \beta\equiv\frac{\mu^2 \Delta z}{p_0^+x(1-x)}. 
\end{equation}

\begin{figure*}[!t]
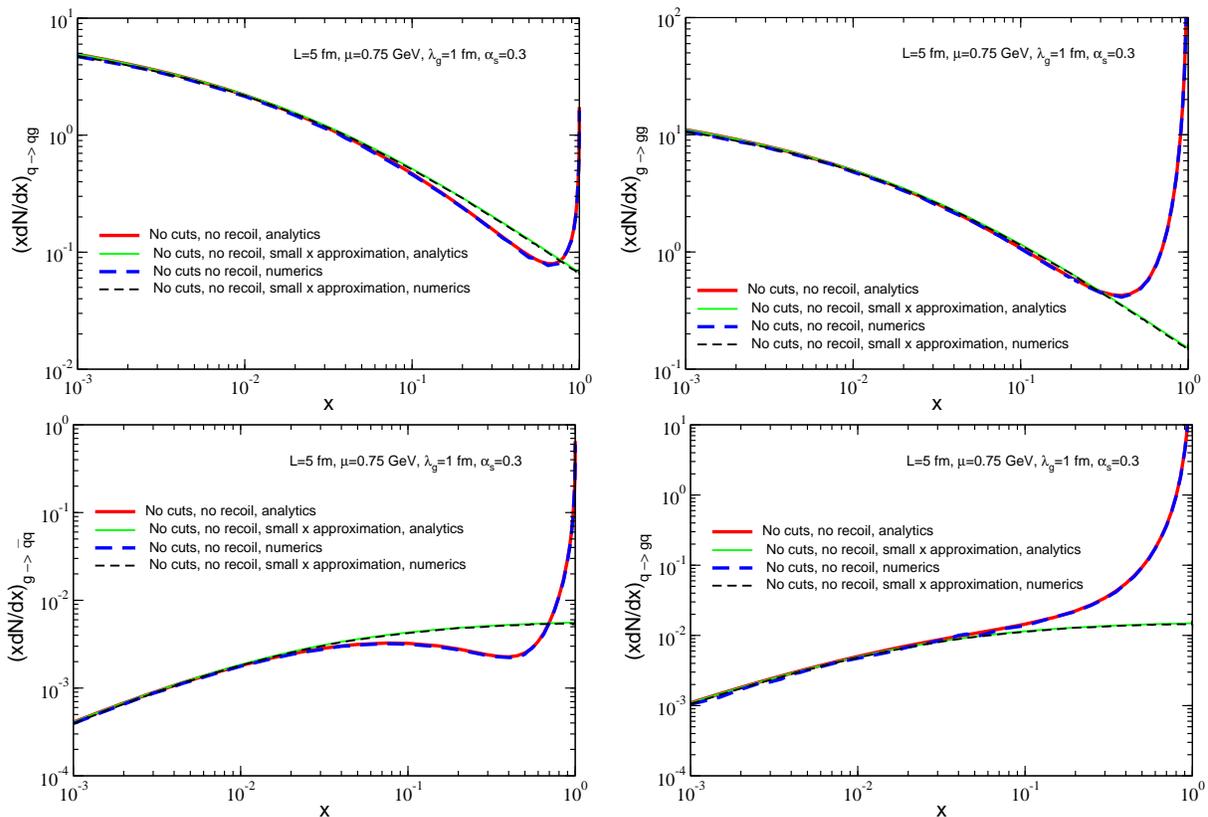

\includegraphics[width= 220pt]{fig1.eps} \hspace*{0.1in}\includegraphics[width= 220pt]{fig2.eps}\\
\includegraphics[width= 220pt]{fig3.eps} \hspace*{0.1in} \includegraphics[width= 220pt]{fig4.eps}
\caption{The intensity spectrum $x(dN/dx)$ for infinite phase space cuts and neglecting nuclear recoil is shown as a function of the splitting parameter $x$.  
Comparison of the analytic formulas in \eq{finalanalytics01}-\eq{finalanalytics03} (solid lines) 
to a numerical integration 
method (dashed lines) are presented. We also illustrate the difference between the full 
in-medium splitting results and the small-$x$ approximation on the example of a parton of
initial energy  $E_0=100\text{GeV}$. The medium parameters are set to: 
$\mu=0.75\GeV, \,\lambda_g=1\text{ fm}, \,L=5\text{ fm}$ for definiteness and the  scattering 
length is independent of $\Delta z$.  }
\label{fig:numerics}
\end{figure*}

In order to perform the remaining $\Delta z$ integral, one has to specify the 
geometry and  an expansion scenario for the QCD medium that the partons traverse.  
Even the simplest realistic model of the medium, which includes the Glauber nuclear 
geometry and Bjorken expansion, requires numerical evaluation of the in-medium splitting 
kernels. One can validate the numerical simulation techniques by comparing the results 
to  closed-form analytic formulas for uniform QCD matter, where the parton mean free paths 
$\lambda_q,\,  \lambda_g$  and trajectory length $L$ are fixed. Thus, it is instructive to have analytical formulas
for splitting intensity for uniform QCD matter. By assuming constant $\lambda_{q, g}$ as a function of $\Delta z$, and integrating expressions in \eq{split1analytics}-\eq{split4analytics} over $0<\Delta z<L$,  we obtain the following differential 
splitting intensities:
\begin{widetext}
\begin{eqnarray}\label{finalanalytics01}
&&x\left(\frac{dN}{dx}\right)^{\infty, \text{static}}_{q\rightarrow qg}=
 x\frac{\alpha_s}{2\pi} C_F \frac{ 1+(1-  x)^2 }{x}
\frac{ L}{\lambda_g} \left( g[\gamma]+g[\gamma(1-x)^2]-\frac{1}{N_c^2}g[\gamma x^2] \right),\\
&&x\left(\frac{dN}{dx}\right)^{\infty, \text{static}}_{g\rightarrow gg}= x\frac{\alpha_s}{2\pi} 2 C_A
\left(\frac{x}{1-x}+\frac{1-x}{x}+x(1-x)\right) \frac{ L}{\lambda_g} 
\left( g[\gamma x^2]+g[\gamma(1-x)^2]+g[\gamma] \right) ,
\label{finalanalytics02}\\
&&x\left(\frac{dN}{dx}\right)^{\infty, \text{static}}_{g\rightarrow q\bar{q}}
=x\frac{\alpha_s}{2\pi} T_R \left(x^2+(1-x)^2\right) \frac{ L}{\lambda_q}  \left[{\frac{2N_c^2}{N_c^2-1}
\left(g[\gamma x^2]+g[\gamma(1-x)^2]\right)-\frac{2}{N_c^2-1}g[\gamma]}\right] ,
\label{finalanalytics03}
\end{eqnarray}
\end{widetext}
where the function $g$ is given by:
\begin{eqnarray}
w \, g[w]&\equiv&  \frac{\pi}{2}(1-\cos(w))+(\gamma_{\text{E}}-1)w+w \ln(w) \nonumber \\
 &&+\text{Si}(w)\cos(w)-\text{Ci}(w)\sin (w) ,
\label{finalanalytics1}
\end{eqnarray}
and $\gamma$ is defined as:
\begin{eqnarray}
\gamma &\equiv& \frac{\mu^2 L}{p^+_0 x(1-x)}.
\end{eqnarray}
The intensity spectrum for the last splitting $q\rightarrow gq$ can be obtained from 
substitution $x\rightarrow 1-x$ in the $q\rightarrow qg$ splitting and is given in 
\eq{split4analytics}.

\section{Numerical results}
\label{numericssec}

In this section we study the effects of kinematic cuts and recoil of the medium by evaluating $dN/dx$ numerically. In so doing, we demonstrate control over 
the numerical evaluation, keeping in mind that future applications will require such 
approach to incorporate the finite kinematics, the spatially non-uniform and time-dependent
density of the QCD matter, and the recoil of the in-medium partons. For each splitting  
we consider the full result given by \eq{CohRadSX1} - \eq{CohRadSX2} and compare it to the 
small-$x$ limit presented in \eq{smallx}. In this paper we consider a medium of uniform 
density for simplicity and set the parameters of the simulation
as follows: the typical inverse range of the parton scattering in the medium is $\mu=0.75\GeV$, 
the size of the QCD medium is $L=5\text{ fm}$, the gluon mean free path in matter is 
$\lambda_g=1\text{ fm}$,  and the parent parton energy is $E_0 = p^+_0/2=100\ \GeV$. 

For infinite limits of the $\vc{k}_{\perp}, \vc{q}_{\perp}$ integrations, ignoring the medium recoil 
effects and assuming static QCD matter, we checked numerically our analytic formulas 
in \eq{finalanalytics01} - \eq{finalanalytics03}. We found perfect agreement that validates 
the numerical integration methods and the analytic results. This can be seen from 
figure~\ref{fig:numerics}.  Solid lines  represents the analytic results of  
\eq{finalanalytics01} -  \eq{finalanalytics03}. Dashed lines represent numerical results. Our conclusions 
are valid for both the full in-medium splitting intensity $x(dN/dx)$ and its small-$x$ limit. Note that 
for such comparison to be possible we have retained the subleading  ${\cal O} (x)$ term for the 
$g \rightarrow q\bar{q}$ and  $q \rightarrow gq$ processes. As expected, the deviation between the 
 full in-medium splittings (red and blue lines)  and their small-$x$ approximation (green and black lines) 
is the largest as $x \rightarrow 1$. For intermediate $x \sim 0.5$ the deviation is on the order of a factor 
of 2 and changes sign.

\begin{figure*}[!t]
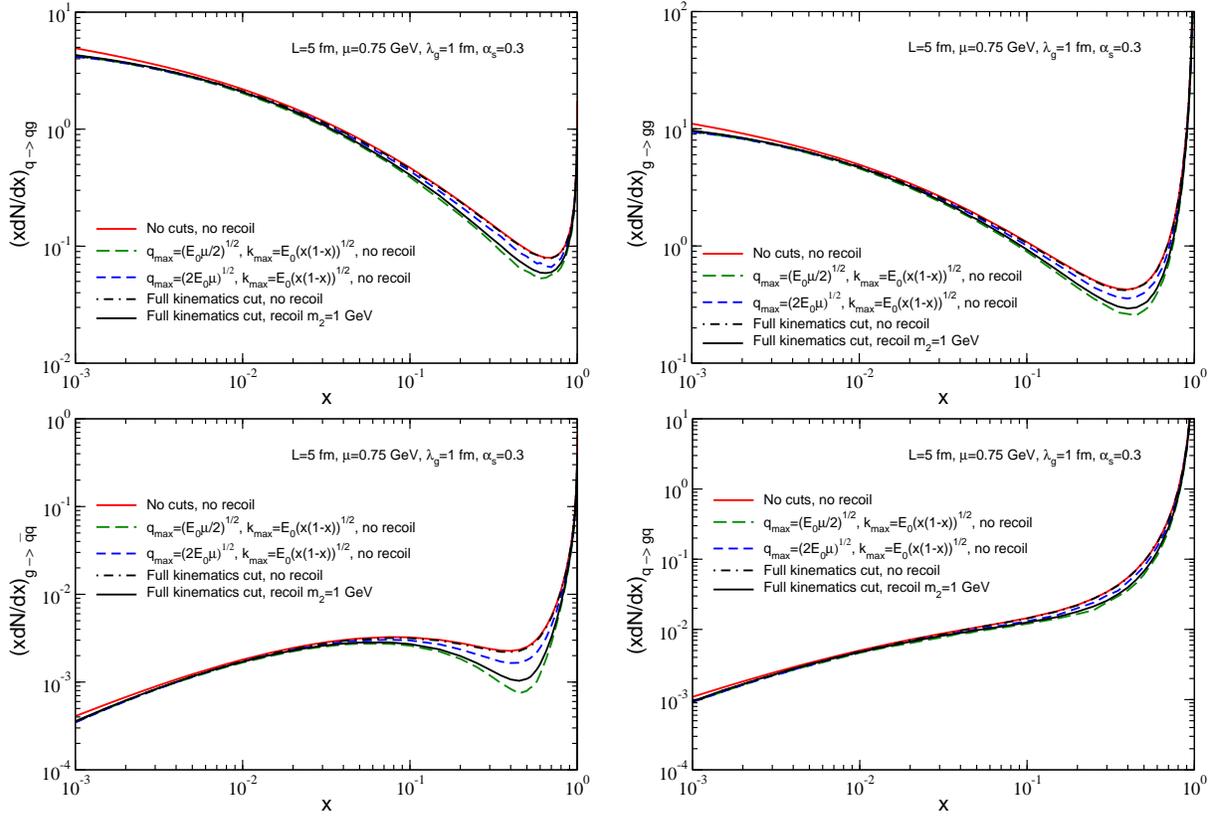

\includegraphics[width= 220pt]{fig5.eps} \hspace*{0.1in}
\includegraphics[width= 220pt]{fig6.eps}\\
\includegraphics[width= 220pt]{fig7.eps} \hspace*{0.1in} 
\includegraphics[width= 220pt]{fig8.eps}
\caption{Illustration of the effect of phase space cuts and medium recoil on the medium-induced parton 
splitting. The same QCD medium parameters and initial jet energy as in figure~\ref{fig:numerics}
are used. }
 \label{fig:numerics2}
\end{figure*}

In figure~\ref{fig:numerics2} we present the comparison of the  splitting intensities  without 
transverse momentum   cuts and without parton recoil in the medium (solid red curve) to  
three different cut 
scenarios. In all three cases we use the same cut on $k_{\text{max}}=\sqrt{Q^2 x(1-x)} $, 
which is unambiguous, and we choose  $Q=E_0$. The three scenarios for the $\vc{q}_{\perp}$ cut are: 
\textit{a}) the dashed green curve corresponds to $q_{\text{max}}=\sqrt{\mu E_0/2}$, \textit{b}) 
the dashed blue curve corresponds to $q_{\text{max}}=\sqrt{2\mu E_0}$, \textit{c}) the
dot-dashed black curve corresponds to the exact  phase space, given by $2\rightarrow2$ 
scattering available phase space. Finally, the solid black curve includes the recoil effect which 
is calculated by substituting the normalized cross section in \eq{infsigmael} by the 
$2\rightarrow 2$ $t$-channel differential cross section, which can be found in Eqs.~(3.2-3.3) in Ref.~\cite{Ovanesyan:2011xy}. 
From the definition of cut on $\vc{k}_{\perp}$, for small $x$ we have $k_{\text{max}}\sim Q\sqrt{x}\rightarrow 0$. From figure~\ref{fig:numerics2} one can see that for small $x$ the cut on $\vc{k}_{\perp}$ is 
the only one that affects the splitting intensity, since all three types of cuts on ${\vc{q}_{\perp}}$ 
give practically identical results. As far as the intermediate $x$ region is concerned, the cut on 
$\vc{k}_{\perp}$ does not play a significant role since from the definition $k_{\text{max}}(x\sim1/2)\sim Q/2$, thus the observable difference must be attributed to the cut on $\vc{q}_{\perp}$ for this region. The kinematic cut on $\vc{q}_{\perp}$, however,
can lead to a factor of 2 variation of the in-medium parton splitting intensities at intermediate $x$.
Note that for intermediate $x$ the third cut on $\vc{q}_{\perp}$,  which corresponds to full kinematics 
but retains the $1/\vc{q}_{\perp}^4$ dependence of the scattering cross section~\eq{infsigmael}, 
agrees perfectly with  the uncut solid red curve. In this case,  cuts alone  
(in the sense of full kinematics) do not affect significantly the in-medium branching 
processes. We find that what affects the splitting intensity is the deviation between  
the exact scattering cross section from~\cite{Ovanesyan:2011xy} and the approximate power-law form
in~\eq{infsigmael}.  This is illustrated in  figure~\ref{fig:numerics2} by the solid black curve 
that pushes the intensity of the medium-induced branching processes down when compared to 
the dot-dashed black curve. We finally note that if one wishes to simplify the calculation and
use the approximate form~\eq{infsigmael} for in-medium parton scattering the most adequate transverse 
momentum cut would be  $q_{\text{max}}=\sqrt{\mu E_0}$. Finally, for $x\rightarrow 1$, we find that $k_{\text{max}}\sim Q\sqrt{1-x}\rightarrow 0$, and all the splitting intensities with phase space cuts turn over at large enough $x$, which is not visible in figure \ref{fig:numerics2} because we do not plot values of $x$ very close to 1.

Numerically, all effects: finite $x$, phase space cuts, recoil effect are of the same order at 
high energies.  In addition, we observe that parton recoil, similar to finite $x$ corrections 
appears at intermediate $x$, while phase space cuts play role both for small $x$ and intermediate $x$.

\section{Conclusions}\label{conclusions}
In this Letter we derived the medium-induced  parton splittings for quarks and gluons produced in large
$Q^2$ scattering processes that subsequently traverse a region of dense QCD matter and 
undergo final-state interactions 
using a recently constructed   effective theory $\text{\SCETG}$~\cite{Ovanesyan:2011xy}. 
Our results include both the contributions from the in-medium parton scattering  and their  dominant 
interference with the vacuum-like branching processes~\cite{Gyulassy:2000er,Vitev:2007ve}. 
These formulas are valid for an arbitrary splitting parameter $x$ and include all leading terms consistent 
with the power counting of $\text{\SCETG}$~\cite{Ovanesyan:2011xy}. Our results for 
the $q\rightarrow qg,\, q\rightarrow gq,\,    g\rightarrow gg, \, g\rightarrow q\bar{q}$,  including 
Landau-Pomeranchuck-Migdal coherence and interference effects,  are presented in  
\eq{CohRadSX1}-\eq{CohRadSX2} and are the main new findings reported in this Letter. 
We verified in \eq{smallx} that in the small-$x$ approximation our formulas 
simplify considerably and reduce to the soft gluon emission results of the Gyulassy-Levai-Vitev approach to 
inelastic parton interactions in dense QCD matter~\cite{Gyulassy:2000er,Vitev:2007ve}.  In this case,
to leading power in $1/x$, only two medium-induced splitting kernels survive that do not 
change the flavor of the leading parton and have a natural interpretation in terms of parton
energy loss.

Neglecting the recoil of the partons in the QCD medium and ignoring the phase space cuts, we 
derived  fully analytic formulas for the in-medium splitting intensity, given in 
\eq{finalanalytics01}-\eq{finalanalytics03}. We see the main utility of these formulas as a 
convenient cross check for the numerical simulations, which we have demonstrated in this Letter. 
Our full results provide a basis for further improvements in the jet quenching  
phenomenology~\cite{Neufeld:2010fj}-\cite{Mironov:2011zz} by including the 
following effects: $a)$ finite $x$ corrections in the $q\rightarrow qg$ and $g\rightarrow gg$ splittings,
consistent with the power counting of \SCETG, 
$b)$ new splittings for $g\rightarrow q\bar{q}$, $q\rightarrow gq$ from coherent final-state
interactions,  $c)$ exact parton recoil kinematics and, $d)$ exact phase space cuts.
All of the above effects can be of the same order, as one can see from our numerical simulation 
results in section~\ref{numericssec}. This program of improving the theoretical accuracy of 
jet quenching simulations is especially interesting in light of recent RHIC and LHC
heavy ion results ~\cite{Aamodt:2010jd}-\cite{Dainese:2011vb}, \cite{Salur:2009vz}-\cite{Chatrchyan:2011ua} and we plan to show first phenomenological 
applications in the near future.

\begin{acknowledgments}
This research is  supported by the US Department of Energy, Office  of Science, under
Contract No. DE-AC52-06NA25396 and in part by the LDRD program at LANL and the JET topical 
collaboration.
\end{acknowledgments}

\end{document}